\documentstyle[amssymb,twoside,fleqn,espcrc2]{article}

\newcommand{\AmS}{{\protect\the\textfont2
  A\kern-.1667em\lower.5ex\hbox{M}\kern-.125emS}}
\def\inprod{\mathop{\kern -0.05em\raise -0.1em\hbox{%
  \vrule height 0.03em width 0.6em depth 0em%
  \vrule height 0.7em width 0.03em depth 0em}\kern 0.1em}\nolimits}
\def\d{\mbox{\sf d}}
\def\Rn{{\Bbb R}^n}
\def\BrX{\breve{{\bf X}}}

\title{Boundary terms and their Hamiltonian dynamics}

\author{V.O. Soloviev\address{Theory Division,
        Institute for High Energy Physics,\\
        142284, Protvino, Moscow region, Russia}}

\begin{document}

\begin{abstract}
It is described
how the standard Poisson bracket formulas should be
modified in order to incorporate integrals of divergences into the
Hamiltonian formalism  and why this is necessary.
Examples from Einstein gravity and Yang-Mills gauge field theory are
given.
\end{abstract}

\maketitle

\section{INTRODUCTION}

Hamiltonian mechanics traditionally serves as an \'etalon part of the
mathematical physics both for physicists and mathematicians \cite{Arn,MR}.
Many
of its geometrical constructions are exported now to the field theory.
For example, the Schouten-Nijenhuis bracket
\cite{Nij}
turned out to be
extremely useful in the search for integrable models
during the last 20 years.
In the pioneering article \cite{GD}
this approach had been called
as the formal variational calculus because in proving
theorems it was possible to replace functionals by functions, even
polynomials.
Presently there are some monographs available
where the method is presented
comprehensively \cite{Olv,Dorf}.

The main restriction
required by this approach is the freedom
to integrate by parts.
This requirement is fulfilled if the fields
are rapidly decaying at spatial infinity (as the massive fields usually do)
or if the periodic boundary conditions
are imposed.
However, these cases
do not cover
all types of the physically interesting boundary
behaviour.
Often one has to deal with a slow decay of the gauge or
gravitational fields at spatial infinity \cite{RT,Sol0}, with
nonperiodical boundary conditions for a
media
moving in a
finite domain \cite{LMMR} or for integrable models
\cite{FT,BFT,APP}.
Can we still use the Hamiltonian approach in such cases?  Of
course, the answer is positive and the cited papers just do that.
What we are trying to do here is to propose a more general framework
for the nontrivial boundary problems.

In this report we intend to show  that the main geometrical concepts
of the formal variational calculus, and, consequently,
the Hamiltonian mechanics, can be preserved without the standard requirement
on the freedom to integrate by parts.
In other words we propose an approach
that does not neglect  surface integrals. This
purpose
can be achieved by introduction of a new grading into the formal
variational calculus \cite{Sol2}
and a new pairing
compatible with this
grading.
First of all, the revision consists in
the modification of
formulas for the Poisson bracket in field theory \cite{Sol1}.

\section{THE POISSON BRACKET}

Briefly then,
we will modify the standard formula for the Poisson bracket
by surface terms in such a way that all its general axiomatic properties
(bilinearity, antisymmetry, Jacobi identity and closeness,
i.e.,
the requirement to remain in a
given space of functionals) \cite{Olv}
will be preserved without discarding any term in the course of integration by
parts.

Being one of the main elements of the Hamiltonian formalism
the Poisson bracket itself is not elementary and
may be considered
as a composite structure

\begin{equation}
\{ F,G\}=\d G\inprod\d F\inprod\Psi.
\end{equation}
Its elements are: the differentials of functionals $\d F$, $\d G$,
the Poisson bivector $\Psi$ and the pairing (or interior product)
operation $\inprod$. It
turns out
that in order to take care
of
all surface integrals we should revise all the three constituents
of the bracket.

There are two ways to write a local functional: as
an integral
of some smooth function $\phi^{(J)}_A(x)$ of the fields and their spatial
derivatives up to some finite order over the prescribed domain $\Omega$ in
$\Rn$, or as the integral over all the space $\Rn$ but with the
characteristic function of the domain $\theta_{\Omega}$ included into the
integrand

\begin{equation}
F=\int_{\Omega}f\left(\phi^{(J)}_A(x)\right)=\int\theta_{\Omega}f.
\end{equation}
Henceforth
we will consider the space $\Rn$, use
the
Einstein rule for
summations and the multi-index notations $J=(j_1,...,j_n)$ where
$j_i\geq 0$

\begin{equation}
\phi_A^{(J)}=
{{\partial^{|J|}\phi_A}\over{\partial^{j_1}x^1...{\partial^{j_n}x^n}}},
\qquad |J|=j_1+...+j_n.
\end{equation}
The reader unfamiliar with these notations may first
have in mind the one-dimensional case, then $J$  simply is the order
of spatial coordinate derivative.
Binomial coefficients for multi-indices are

\begin{equation}
{J \choose K}={j_1\choose k_1}\cdots{j_n\choose k_n},
\end{equation}
\begin{equation}
{j \choose k}= \cases {
j!/(k!(j-k)!) & if  $ 0\le k \le j$; \cr
0  & otherwise. \cr }
\end{equation}
With the help of them we introduce
the so-called higher Eulerian operators \cite{KMGZ,Ald,Olv}

\begin{equation}
E^J_A(f)=(-1)^{|K|+|J|}{K\choose
J}D_{K-J}{{\partial f} \over{\partial\phi_A^{(K)}}},\label{eq:he}
\end{equation}
where

\begin{equation}
D_i={{\partial}\over{\partial x^i}}+\phi_A^{(J+i)}{{\partial}\over
{\partial\phi_A^{(J)}}}, \quad D_J=D_1^{j_1}...D_n^{j_n}.
\end{equation}
In the framework of the
standard
approach the differential of
a
local functional
is given by the Euler-Lagrange derivative $E^0_A(f)$

\begin{equation}
\d F=\int_{\Omega}E^0_A(f)\delta\phi_A.
\end{equation}
This is fully justified
if
 all variations $\delta\phi_A$ and all
their spatial derivatives are zero on the boundary. In
a
 more general
case $E^0_A(f)$ gives us only a part of the full variation. As a
consequence of
that
fact Euler-Lagrange derivatives may not commute
\cite{And}.  In
 turn this leads us to the conclusion \cite{Sol1.5} that
transformations of the form

\begin{equation} q^A(x)\to q^A(x),\quad p_A(x)\to
p_A(x)+{{\delta F[q]}\over{\delta q^A(x)}}, \label{eq:ash_trans}
\end{equation}
are, generally speaking,
 canonical only up to surface terms.
Finally,
as
the standard proof of the Jacobi identity in mechanics is based on
the commutativity of the mixed second derivatives, in field theory
the Jacobi identity for
 functionals may not be true even
if the
fields themselves have
 the canonical Poisson brackets.

To improve the situation we allow arbitrary variations on the boundary
\begin{eqnarray}
\d F & = & \int{{\delta F}\over
{\delta\phi_A}}\delta\phi_A\equiv
\int_{\Omega}f'_A(\delta\phi_A)\equiv\nonumber\\
& \equiv & \int_{\Omega}
D_J\left(E^J_A(f)\delta\phi_A\right),
\end{eqnarray}
where the differential is written consequently
by using the full variational derivative \cite{Sol1}

\begin{equation}
{{\delta F} \over {\delta\phi_A}}= (-1)^{|J|}E^J_A(f)
\theta^{(J)}_{\Omega}\equiv E^0_A(\theta_{\Omega}f),\label{eq:varfull}
\end{equation}
the Frech\'et derivative

\begin{equation}
f'_A={{\partial f}\over{\partial\phi_A^{(J)}}}D_J,
\end{equation}
and the higher Eulerian operators (\ref{eq:he}).

The
second constituent, a Poisson bivector, is given, loosely speaking, by
Poisson brackets of the fields. These brackets are called local if
they are proportional to the $\delta$-function and a finite number of its
derivatives (ultralocal, if derivatives are absent)

\begin{equation}
\{\phi_A(x),\phi_B(y)\}=\hat I_{AB}(x)\delta(x,y),
\end{equation}
where

\begin{equation}
\hat I_{AB}(x)=I^L_{AB}D_L, \qquad I^L_{AB}=I^L_{AB}(\phi_C^{(J)}).
\end{equation}
The new feature of our formalism is
that the
surface contributions to
this brackets  are allowed \cite{Sol2}

\begin{equation}
\{\phi_A(x),\phi_B(y)\}= \theta^{(K)}_{\Omega}(x)\hat I^{\langle K\rangle}_
{AB}(x)\delta(x,y).
\end{equation}
For example,
Ashtekar's transformation in the canonical gravity
which is of the type (\ref{eq:ash_trans}) leads to
the generalized form
of the Poisson brackets given above \cite{Sol1.5}.
This is a rather general feature of transformations of this type.
 We consider another
example connected with the nonlinear Schr\"odinger equation in other
place \cite{Sol3}.

Even if the Poisson brackets of fields do not contain surface contributions
such contributions may arise in the calculations of the Poisson algebras
for some transformation generators constructed by means of these fields.
It is so because these nonstandard terms may be a result of moving the
derivatives of the $\delta$-function from one of its arguments to another.
The standard rule

\begin{equation}
\hat I_{AB}(x)\delta(x,y)=\hat I_{AB}^{\ast}(y)\delta(x,y),
\end{equation}
is applicable but with the definition of the adjoint operator modified
to preserve all the boundary terms
\begin{eqnarray}
I^{\ast\langle J\rangle
M}_{AB}&=&
(-1)^{|K|}{K\choose L}{K-L\choose M}\times\\
&\times&D_{K-L-M}I^{\langle J-L\rangle K}_{BA}.\nonumber
\end{eqnarray}
For example, if we preserve the boundary contributions, then
the usual formula

\begin{equation}
\left({{\partial}\over{\partial
x^i}}+{{\partial}\over{\partial y^i}}\right)\delta(x,y)=0,\label{eq:usual}
\end{equation}
should be replaced by the new one

\[
\left(\theta_{\Omega}(x){{\partial}\over{\partial
x^i}}+\theta_{\Omega}(y){{\partial}\over{\partial y^i}}\right)\delta(x,y)=
-\theta^{(i)}_{\Omega}\delta(x,y).
\]
This solves
one
paradox which arises in understanding the result
obtained for asymptotically flat spaces in
the
canonical General Relativity.
It is shown \cite{RT,Sol0} that

\begin{equation}
\{ H(\xi),H(\eta)\} \approx H([\xi,\eta]),\label{eq:alg}
\end{equation}
where $\xi^{\alpha}$, $\eta^{\beta}$ are
the
 Killing vectors of the background
flat metric, $[\xi,\eta]$ is their Lie bracket and $H(\xi)$, $H(\eta)$,
$H([\xi,\eta])$ are generators with nonzero surface terms. The paradox
is in the observation that if we first consider integrands of (\ref{eq:alg})
and calculate their Poisson brackets according to (\ref{eq:usual}) then we
will get zero result due to the closed constraint algebra of
the General Relativity.

The third step in our revision of the Poisson bracket formula is dictated
by purely mathematical reasons.
Speaking in mathematical terms,
the extension of the formal
variational calculus proposed above is the introduction of a new grading.

A  grading in linear space $L$ is a decomposition of it into
a
 direct sum
of subspaces, with a special value of some function $p$ (grading function)
assigned to all the elements of any subspace \cite{Dorf}.

Here the function $p$ takes its values in the set of all positive
multi-indices and
thus,

\begin{equation}
L=\bigoplus\limits_{J=0}^{\infty} L^{\langle J\rangle}.
\end{equation}
Elements of each subspace are called homogeneous.

A bilinear operation $x,y\mapsto x\circ y$, defined on $L$, is said to be
compatible with the grading if the product of any homogeneous elements
is also homogeneous, and if

\begin{equation}
p(x\circ y)=p(x)+p(y).
\end{equation}
It is necessary to define the pairing between 1-forms and 1-vectors, which
then really will induce all other operations,
as an operation compatible with the introduced grading. In our
notions 1-forms are

\begin{equation}
\alpha=\int \theta^{(J)}_{\Omega}\alpha_{AK}^{\langle J\rangle}
D_K\delta\phi_A,
\end{equation}
whereas 1-vectors are

\begin{equation}
\psi=\int \theta^{(I)}_{\Omega}\psi_{BL}^{\langle I\rangle}
D_L\biggl({{\delta}\over{\delta\phi_B}}\biggr).
\end{equation}
Their bases are dual

\begin{equation}
\left\langle\delta\phi_A(x),{{\delta}\over{\delta\phi_B(y)}}\right\rangle
=\delta_{AB}\delta(x,y),
\end{equation}
and the graded differential operators
\begin{eqnarray}
\hat\alpha&=& \theta^{(J)}_{\Omega}\alpha_{AK}^{\langle J\rangle}
D_K,         \\
\hat\psi&=& \theta^{(I)}_{\Omega}\psi_{BL}^{\langle I\rangle}
D_L,
\end{eqnarray}
serve as coefficients of the decomposition over these bases. Let us define
a trace  as

\begin{equation}
{\rm Tr}(\hat\alpha\hat\psi)=\theta^{(I+J)}_{\Omega} D_L\alpha_{AK}^{\langle
J\rangle} D_K\psi_{AL}^{\langle I\rangle},
\end{equation}
%which %
that
is evidently a bilinear and commutative operation.
It is also easy to check that

\begin{equation}
 {\rm Tr}((D\hat\alpha)\hat\psi)=D{\rm
Tr}(\hat\alpha\hat\psi).
\end{equation}
This important property of the trace operation
allows us to use it for the definition of the pairing

\begin{equation}
\alpha(\psi)=\psi\inprod\alpha=\int {\rm Tr}(\hat\alpha\hat\psi),
\end{equation}
which is independent on the ambiguity in the representations of the
operators $\hat\alpha$, $\hat\psi$
 following from the freedom to do the formal integration by parts.
For example, we can remove all the derivatives from the basis elements
\begin{eqnarray}
\alpha=\int \theta^{(J)}_{\Omega}\tilde\alpha_{A}^{\langle J\rangle}
\delta\phi_A,\\
\psi=\int \theta^{(I)}_{\Omega}\tilde\psi_{B}^{\langle I\rangle}
{{\delta}\over{\delta\phi_B}},
\end{eqnarray}
thus transforming $\alpha$ and $\psi$ to the so-called canonical form (compare
with \cite{Olv}).  This  formal integration by parts does not change any
integral over the finite domain $\Omega$ and is useful for the illustration
of analogy with the standard formalism.  To return back to the usual formal
variational calculus we should only put $\theta_{\Omega}(x)
\equiv 1$, i.e., $\Omega=\Rn$ then the
``columns'' like $\alpha^{\langle J\rangle}$ will be reduced to their first
terms $\alpha^{\langle 0\rangle}$ and for the canonical representation of the
1-vector and 1-form their pairing will be reduced to the standard one

\begin{equation}
\psi\inprod\alpha=\int
\tilde\alpha_A^{\langle 0\rangle} \tilde\psi_A^{\langle 0\rangle}.
\end{equation}
%So, the pairing introduced above is compatible with the proposed grading.

After making the above three steps: the revision of differential,
Poisson bivector and  pairing we can obtain the new formula. There are
at least three ways to write it, in correspondence to
 the three ways to write
the differential of a local functional:
through the full variational derivatives

\begin{equation}
\{ F,G\} =\int\int {{\delta F} \over {\delta \phi_A(x)}}
 {{\delta G} \over {\delta \phi_B(y)}}\hat I_{AB}(x)\delta(x,y),
\end{equation}
through the Frech\'et derivatives

\begin {equation}
\{ F,G \} =
\int\theta^{(J)}_{\Omega} {\rm Tr}\left( f'_A\hat I^{\langle J\rangle}_{AB}
g'_B \right), \label{eq:brack1}
\end{equation}
and through the higher Eulerian operators

\begin{equation}
\{ F,G \} =
\int\theta^{(J)}_{\Omega} D_{P+Q}\left( E^P_A(f)\hat I^{\langle
J\rangle}_{AB} E^Q_B(g) \right).\label{eq:brack2}
\end{equation}

\section{EXAMPLES}

The calculation of the Poisson brackets by the new formulas can be made
in not more complicated way
than by the old ones. First, to get the Frech\'et
derivative from the first variation is even easier than to get the standard
Euler-Lagrange derivative because the integration by parts is not needed.
Second, we can exploit covariance properties and use the covariant
derivatives instead of the ordinary ones.

As an example, we calculate the Poisson brackets of the two spatial
diffeomorphism generators in the canonical General Relativity for a finite
domain
\begin{eqnarray}
H(N^i)&=&\int_{\Omega}\pi^{ij}(\nabla_jN_i+\nabla_iN_j)d^3x=\\
&=&\oint_{\partial\Omega}2\pi_i^jN^idS_j-\int_{\Omega}2N^i\nabla_j
\pi_i^jd^3x.
\nonumber
\end{eqnarray}
  From the first variation we get
\begin{eqnarray}
h'_{\pi^{ij}}(N^i)&=&\nabla_jN_i+\nabla_iN_j,\\
h'_{\gamma_{ij}}(N^i)&=&\pi^{ik}
\nabla_kN^j+\pi^{kj}\nabla_kN^i+N^k\pi^{ij}\nabla_k,\nonumber
\end{eqnarray}
and then calculate the Poisson bracket according to formula
(\ref{eq:brack1})

\begin{equation}
\{H(N^i),H(M^i)\}= H({[N,M]}^i),
\end{equation}
where

\begin{equation}
{[N,M]}^i=N^kM^i_{\ ,k}-M^kN^i_{\ ,k}.
\end{equation}
Let us mention that the calculation according to the standard formula
gives additional surface contribution violating the diffeomorphism algebra
\begin{eqnarray}
\Delta H&=&\oint_{\partial\Omega}\pi^{ij}\left((\nabla_jN_i+\nabla_iN_j)M^k-
\right.\nonumber\\
&-&\left. (\nabla_jM_i+\nabla_iM_j)N^k\right)dS_k.
\end{eqnarray}
This violation is zero in the only case when $N^i$ and $M^i$ are the Killing
vectors on the boundary. Therefore we have  a free boundary closure of
the spatial diffeomorphism algebra by means of the new brackets.

As a second example let us consider the Yang-Mills field in a finite
domain.  It is suitable to use orthonormal curvilinear coordinates
$X_k$ in  ${\Bbb R}^3$, so that they are compatible with the boundary
$\partial\Omega=\{X_k: X_1=R={\rm const}, \BrX=X_2, X_3\}$. If $x_k$ are
Cartesian coordinates then the local frame

\begin{equation}
e^{(k)}_i=h_k^{-1}{{\partial x_i}\over{\partial X_k}}, \qquad
h_k=\sqrt{\left({{\partial x_i}\over{\partial X_k}}\right)^2},
\end{equation}
can be used.
The Hamiltonian has
the
 form \cite{Tim}
\begin{eqnarray}
H_{\Omega} &=& \int_{\Omega} d{\bf x}\ \Biggl(\Biggr. \frac{1}{2}
(E^{a}_{i})^{2} + \frac{1}{4} (F^{a}_{ij})^{2} \nonumber \\
&-& A_{0}^{a} \,\bigl( \partial_{i}E^{a}_{i} - g t^{abc}A^{b}_{i}
E^{c}_{i} \bigr) \Biggl.\Biggr).
\end{eqnarray}
It should be  accompanied by
the
 surface contribution

\begin{equation}
\Delta H_{\Omega} = \int_{\partial V} d\BrX\ A^{a}_{0}
\left( \frac{h}{h_{1}} E^{a}_{(1)} + \chi^{a} \right),
\end{equation}
where $h=h_1h_2h_3$, and $\chi^{a}$ are the surface
variables
possessing
the Poisson bracket

\begin{equation}
\{\chi^{a}(\BrX),\chi^{b}(\BrX')\}=
gt^{abc}\,\chi^{c}(\BrX)\,\delta(\BrX,\BrX'),
\end{equation}
and
commuting
 with the volume variables. After
fixing
 the spatial Fock-Schwinger
gauge, $A^a_{(1)}=0$,
the Hamiltonian evolution on the boundary is
given by

\begin{eqnarray}
\dot{E}_{(1)}^{a}(R,\BrX)& =& - gt^{abc}E^{b}_{(1)}(R,\BrX)
A^{c}_{0}(R,\BrX),\\
\dot{\chi}^{a}(\BrX)&=&gt^{abc}\,\chi^{b}(\BrX)\,A^{c}_{0}(R,\BrX),
\end{eqnarray}
where  the boundary condition
$G_{(1)}^a(R,\breve{{\bf X}})=0,$
$G_{i}^a \equiv \nabla_j F_{ij}^a$
compatible with the localized time evolution is also assumed.

Here
the Gauss law constraint is prolonged onto the boundary by introduction
of the surface variables $\chi^{a}(\BrX)$.
The standard approach requires
a fixation of the static
boundary conditions for  ${E}_{(1)}^{a}$, and in its
turn this requires that the Lagrangian multiplier $A^{c}_{0}$ should
be zero on the boundary. Then the boundary conditions are to be
gauge-dependent or $E^a_{(1)}$ be zero. The approach based on
the dynamical boundary
conditions permits to save  gauge invariance on the boundary.

The residual
gauge invariance of the theory is manifested in the above dynamics on the
boundary. This implies that the boundary conditions put onto $E^a_{(1)}$ may
not necessarily be arbitrary to preserve the gauge invariance. It has been
argued that the dependence of the partition function on these boundary
conditions may be considered as a confinement criterion in the SU(N) gauge
theory \cite{Timdiss}, and that the surface terms play an important role for
understanding that phenomenon \cite{ST}.

\section{CONCLUSION}

As we have seen from above, the Poisson structure can be introduced
prior to any boundary conditions. This is analogous to
the  Hamiltonian
mechanics where constraints are treated
later than the Poisson
brackets. We may expect that the treatment of the boundary conditions
could proceed similarly, so that primary and secondary, first and second
class boundary conditions may arise. We may get an analog of the Dirac
bracket at the end of the standard reduction procedure.

What are the Hamiltonian equations generated by the new
bracket? They can be called as a weak form of the equations of motion
\cite{MR}.
If we try to understand $\theta_{\Omega}$-functions
as distributions seriously, then
we will have singular boundary terms in  the equations and may
encounter with ambiguities in solving such equations.  Therefore, the
construction of the closed Poisson algebras with surface terms seems
to be a more promising direction. There we deal with local functionals,
rather than functions, and the Poisson bracket does not
move us out of that class.  It is quite possible that the Hamiltonian
dynamics for the functionals may become of more importance for the
quantum field theory than for the classical one.

\section{ACKNOWLEDGEMENTS}

The present approach appeared partially due to the influence of works
by G.~Morchio and F.~Strocchi on the variables at infinity
\cite{MS} and their
development by N.~Sveshnikov and E.~Timoshenko \cite{ST}.
The latter author has
made an attempt to reproduce the results previously obtained
using the methods of the variables
at infinity with the methods presented here \cite{Tim}.
The most complete presentation of these results at the moment,
unfortunately, exists only in Russian \cite{Timdiss}.
The author gratefully acknowledge numerous productive
discussions with
E.~Timoshenko and a very useful discussion with G.~Morchio.

It is a great pleasure to thank organizers of the Symposium and especially
G.~Weigt, T.~Mohaupt and D.~L\"ust for the invitation, support and kind
hospitality in Buckow and Zeuthen.

\end{document}